\documentstyle[preprint,aps,epsf2]{revtex}

\epsfclipon

\tightenlines

\def\be{\begin{equation}}
\def\ee{\end{equation}}
\def\ba{\begin{eqnarray}}
\def\ea{\end{eqnarray}}

\newcommand{\met}{\hbox{{$E_T$}\kern-1.1em\hbox{/}}\kern+0.55em}

\begin{document}

\preprint{\vbox{
\hbox{OSU-HEP-98-4}
\hbox{OITS-654}
\hbox{FERMILAB-Pub-98/222-T}}}

\title{\vspace{3 cm}Gauge Mediated Supersymmetry Signals at the Tevatron
Involving $\tau$ Leptons}

\author{B. Dutta$^{1,a} $, D.J. Muller$^{2} $ and S. Nandi$^{2,a} $}
\address{$^{1}$Institute of Theoretical Physics, University of Oregon, Eugene, 
OR 97403}
\address{$^{2}$Department of Physics, Oklahoma State University, Stillwater,
OK 74078}
\address{$^{a}$  Fermilab, P.O. Box 500, Batavia, IL 60510 }
\date{\today}
\maketitle

\begin{abstract}
We consider the phenomenology of GMSB models where the lighter stau is
the next to lightest supersymmetric particle. In this situation the 
dominant signals for supersymmetry at the Tevatron are events where
two or three high $p_T$ $\tau$ leptons accompanied by large missing 
transverse energy are produced. This leads to signatures that are
very different from the photonic signals for GMSB 
(where the lightest neutralino is the NLSP) and the dilepton and trilepton 
signals in the usual supergravity models (involving $e$ and/or $\mu$ only)
that have been investigated extensively. We find that the inclusive 
2 $\tau$-jet signature
could be observable at the Tevatron Run II, while the inclusive 
3 $\tau$-jet signature
could be important at Run III.
\end{abstract}


\newpage

\section{Introduction}

Supersymmetry (SUSY) has been the focus of a great deal of experimental 
effort
due to the requirement that the superparticles have masses of
{$\cal O$}(1) TeV or less in order to solve the hierarchy problem. 
Thus there exists the possibility of producing SUSY particles at present 
and the next generation of colliders.
The phenomonelogy of such production depends on the nature of the
supersymmetry breaking.
In typical theories of SUSY breaking, the effects of this breaking,
which occurs in a ``hidden sector", are communicated to the
``visible sector" (which includes the usual particles and their
superpartners) by a ``messenger sector".
Searches for SUSY have mostly been inspired by gravity
mediated SUSY breaking theories. In these theories, the lightest
neutralino is usually the lightest
suspersymmetric particle (LSP).
If R-parity is conserved, the LSP is stable and the decay chains of all
other SUSY particles must eventually produce it.
The LSP leaves the detector undetected thereby making large missing 
transverse energy (\met) an important part of the signature for SUSY.
In spite of extensive experimental searches, 
so far no experimental evidence for SUSY has been found 
at the Tevatron \cite{cdfD03} or at LEP \cite{LEP} except for one
possible $e^+ e^- \gamma \gamma$ plus \met\ event at the 
Tevatron \cite{sp}.

Recently, gauge mediated SUSY breaking (GMSB) models have become very 
popular \cite{{dn},{dmn},{dwt}}. 
The defining characteristic of GMSB models is
that the messenger particles interact with visible sector particles via
gauge interactions.
In these theories, the gravitino is the LSP\@. Most phenomenological studies
and experimental
searches that have used GMSB as a framework have taken  the 
next to lightest SUSY particle (NLSP) to be the lightest neutralino.
When this is the case, the $\chi_1^0$ decays to a photon and a gravitino
($\tilde G$). If this decay takes place within the detector, the signal 
involves high $p_T$ photons accompanied by
large \met\ \cite{dwr}.
For much of the parameter space, however, the lighter of the 
two scalar staus is the NLSP\@. 
In this case, the decays of SUSY particles produce the $\tilde\tau_1$
which subsequently decays to a $\tau$ and a gravitino. 
If the $\tilde\tau_1$ decays occur within the detector, 
signatures for SUSY production will then generally include $\tau$
leptons from the $\tilde\tau_1$ decays and \met\ due to the stable
gravitinos and neutrinos leaving the detector.

It was proposed \cite{nb} 
that GMSB
models where the $\tilde \tau_1$ is the NLSP can lead to unusual
and distinguishing signatures for gaugino production. 
At the Tevatron, these signals arise from chargino pair production 
($\chi_1^+ \chi_1^-$) and from the production of the chargino
with the second neutralino ($\chi_1^\pm \chi_2^0$). The subsequent decays
involve multiple high $p_T$ $\tau$ leptons and possibly substantial \met.
The purpose of this paper is to analyze in detail the signals for these
decay modes at the Tevatron. In particular, we seek to determine the
production rates for various distinguishing final states, the $E_T$ 
spectrum of the $\tau$ jets, and the \met\  distribution for the events.

\section{Mass Spectrum and Production Mechanisms}        

Since the observed signal depends on the masses of the sparticles, we
first begin by describing the model and the corresponding mass spectrum.
In our model, the messenger sector consists of some number of multiplets
that are
$\bar 5 + 5$ representations of SU(5). They couple to a chiral superfield
$S$ in the hidden sector whose scalar component has a vacuum expectation
value (VEV) $\langle s \rangle$ and whose auxiliary component has a VEV
$\langle F_s \rangle$.
By imposing the requirement that the electroweak (EW) symmetry is broken
radiatively, the
particle spectrum and the mixing angles depend on five parameters:
$M$, $\Lambda$, $n$, $\tan \beta$, and the sign of $\mu$. $M$ is the
messenger scale. $\Lambda$ is equal to 
$\langle F_s \rangle$/$\langle s \rangle$ and is related to the SUSY
breaking scale.
The parameter $n$ is dictated by the choice
of the vector-like messenger sector and can take the values 1, 2, 3, or 4 
to satisfy the perturbative unification constraint. 
The definition of $\tan \beta$ is taken as $\tan \beta \equiv v_2/v_1$ where
$v_2$ is the VEV for the up-type ($H_u$) Higgs doublet and $v_1$ is the 
VEV for the down-type ($H_d$) Higgs doublet. The parameter $\mu$ is the 
coefficient in the bilinear term, $\mu H_u H_d$, in the superpotential.
Constraints coming from $b \rightarrow s \gamma$ 
strongly favor negative values for $\mu$ \cite{ddo}
and, in the cases considered in this work, 
$\mu$ is taken to be negative.
Demanding that the EW symmetry be broken radiatively fixes the magnitude
of $\mu$ and the parameter $B$ (from the $B \mu H_u H_d$ term in the 
scalar potential) in terms of the other parameters of the theory. 

The soft
SUSY breaking gaugino and scalar masses at the messenger scale are 
given by \cite{{dn},{spm}}
\be \label{gmass}
\tilde M_i(M) = n \, g(\frac{\Lambda}{M}) \, \frac{\alpha_i(M)}{4 \pi} \, 
\Lambda
\ee
and
\be \label{smass}
\tilde m^2(M) = 2\, n \, f(\frac{\Lambda}{M}) \, \sum^3_{i = 1} \, k_i \, 
C_i \, \left (\frac{\alpha_i(M)}{4 \pi} \right )^2 \, \Lambda^2
\ee                
where the $\alpha_i$ are the three SM gauge couplings and $k_i =$ 1, 1, 
and 3/5
for SU(3), SU(2), and U(1), respectively. The $C_i$ are zero for gauge
singlets and are 4/3, 3/4, and ($Y$/2)$^2$ for the fundamental 
representations of SU(3), SU(2), and U(1), respectively (with $Y$ given
by $Q = I_3 + Y/2$). $g(x)$ and $f(x)$ are messenger scale threshold
functions. We calculate the sparticle masses at the scale $M$ 
using Eqs.~(\ref{gmass}) and (\ref{smass}) and run these 
to the electroweak scale using the appropriate RGE's\cite{bbo}. 
$\mu^2$ is calculated by minimizing the 1 loop corrected Higgs 
potential \cite{bbo}.

The decay chain and hence the signature for the events depends on the
particles initially produced as well as the hierarchy of the masses.
Given the current lower bounds on squark and gluino masses, the production
of strongly interacting sparticles is probably not a viable search modes
for SUSY at the Tevatron Run II\@.
A more likely mechanism for producing SUSY particles is
via EW gaugino production. At the tevatron, chargino pair 
($\chi_1^+ \chi_1^-$) production takes place through s-channel $Z$ and 
$\gamma$ exchange and $\chi_2^0 \chi_1^\pm$ production is through s-channel
$W$ exchange. Squark exchange via the t-channel also contributes to both
processes, but the contributions are expected to be negligible since the 
squark masses are large in GMSB models. The production of 
$\chi^0_1\chi^{\pm}_1$ is suppressed due to the smallness of the coupling 
involved.

Since SUSY breaking is communicated to the visible sector by gauge 
interactions, the mass differences between the superparticles depend
on the their gauge interactions. This creates a hierarchy in mass between
electroweak and strongly interacting sparticles. Eq.\,(\ref{gmass}) shows
that the gluino is more massive than charginos and neutralinos,
while Eq.\,(\ref{smass}) shows that squarks are considerably more 
massive than sleptons.
Given this hierarchy of sparticle masses, there are roughly four possible
cases to consider for EW gaugino production:
\begin{description}
\centering
\item[Case 1:] $m_{\tilde \nu} > M_{\chi_2^0} \agt M_{\chi_1^\pm}
> m_{\tilde e_1, \tilde \mu_1} > M_{\chi_1^0} > m_{\tilde \tau_1}$
\item[Case 2:] $M_{\chi_2^0} \agt M_{\chi_1^\pm} > m_{\tilde \nu}
> M_{\chi_1^0} > m_{\tilde e_1, \tilde \mu_1} > m_{\tilde \tau_1}$
\item[Case 3:] $M_{\chi_2^0} \agt M_{\chi_1^\pm} > m_{\tilde \nu}
> m_{\tilde e_1, \tilde \mu_1} > M_{\chi_1^0} > m_{\tilde \tau_1}$
\item[Case 4:] $m_{\tilde \nu} > M_{\chi_2^0} \agt M_{\chi_1^\pm} 
> M_{\chi_1^0} > m_{\tilde e_1, \tilde \mu_1} > m_{\tilde \tau_1}$
\end{description}
The three sneutrino masses are nearly the same. The lighter of the selectrons
and smuons are essentially right handed and have the same mass. Also, for all
the parameter points we considered, $\chi^{\pm}_1$ and $\chi^0_2$ 
are nearly degenerate.

The possible final states configurations at the Tevatron depend on the
sparticle spectrum, but they will have certain aspects in common. Since
the $\tilde\tau_1$ is the NLSP, 
the various possible decays modes will (usually) produce at least 
two $\tau$ leptons arising from the decays of the $\tilde\tau_1$'s.
In addition, there can also be large \met\ due to the
stable gravitinos and neutrinos escaping detection.

\section{Analysis and Results}

We now give a detailed analysis of the possible Tevatron signatures for
gaugino production in the context of GMSB models where the lightest stau
is the NLSP\@. As mentioned above, the production of SUSY particles in these
models leads to the production of copious quantities of $\tau$ leptons.
Since the lightest chargino is mostly wino, it couples mainly to
left-handed sfermions. Thus, for the examples considered here, 
the dominant decay mode of the chargino is typically 
$\chi_1^\pm \rightarrow \tilde\tau_1 \nu_\tau$ due to the significant
mixing of the left and right handed staus and the 
low mass of the $\tilde\tau_1$.
With the subsequent decay $\tilde\tau_1 \rightarrow \tau \tilde G$, the
expectation is that there are typically two $\tau$ leptons produced in
chargino pair production. Similarly, 
$\chi_2^0 \rightarrow \tilde\tau_1 \tau$ is typically the dominant 
decay mode of the second lightest neutralino. In $\chi_1^\pm \chi_2^0$
production we therefore expect three $\tau$ leptons: one directly from
the $\chi_2^0$ decay and the other from the $\tilde\tau_1$ decays.
So in combined $\chi_1^+ \chi_1^-$/$\chi_1^\pm \chi_2^0$ production,
the expectation is that there will be a significant number of events 
with two and three $\tau$ leptons. 
$\tau$ leptons are identified by their hadronic
decays to thin jets; 
thus we are interested in the probabilities for obtaining
final states with particular numbers of $\tau$-jets.

This analysis is performed in the context of the Main Injector (MI)
and TeV33 upgrades of the Tevatron collider. The center of mass energy
is taken to be $\sqrt{s} = 2$\,TeV and the integrated luminosity is
taken to be 2\,fb$^{-1}$ for the MI upgrade and 30\,fb$^{-1}$ for the 
TeV33 upgrade \cite{TeV33}.

In performing this analysis,
the cuts employed are that final state charged leptons must have 
$p_T > 10$\,GeV and a 
pseudorapidity, $\eta \equiv -\ln ( \tan \frac{\theta}{2} )$ (where 
$\theta$ is the polar angle with respect to the proton beam direction),
of magnitude less than 1.
Jets must have $E_T > 10$\,GeV and $|\eta| < 2$.
In addition, hadronic final states within a cone
size of $\Delta R \equiv \sqrt{ (\Delta \phi)^2 + (\Delta \eta)^2 } = 0.4$
are merged to a single jet. Leptons within this cone radius of a jet are
discounted. 
For a $\tau$-jet to be counted as such, it must have $|\eta| < 1$. 
The most energetic $\tau$ jet is required to have $E_T > 20$\,GeV\@. 
In addition, a missing transverse energy cut of \met\ $> 30$\,GeV is imposed.

We consider each mass case in turn. 
In our analysis, we restrict ourselves to those regions of the 
parameter space 
where the $\tilde\tau_1$ decays promptly to a $\tau$ and a gravitino.
The parameter space is also restricted to those regions
where $m_{\tilde\tau_1} \agt 70$ GeV\@. 
Ongoing LEPII analyses are expected to 
establish this bound soon \cite{pc}. 
With this restriction, we did not find any examples for
Cases 3 and 4.

\subsection{Case 1: $m_{\tilde \nu} > M_{\chi_2^0} \approx
M_{\chi_1^\pm} > m_{\tilde e_1, \tilde \mu_1} > M_{\chi_1^0} > 
m_{\tilde \tau_1}$}

We consider three examples of this case; the masses 
and branching ratios of which are given 
in Table \ref{mass1}. We first consider Example~1 which has 
$\tan \beta = 20$, $\Lambda = 32$\,TeV, $M = 480$\,TeV, and $n = 2$.
Chargino pair production is particularly simple for this case as
$\chi_1^\pm \rightarrow \tilde\tau_1 \nu_\tau$ is not only the dominant
decay mode, but is essentially the only decay mode. Thus in chargino
pair production, two $\tau$ leptons are always produced. 
$\chi_1^\pm \chi_2^0$ production is almost as simple. The main decay
mode of the second heaviest neutralino is 
$\chi_2^0 \rightarrow \tilde\tau_1 \tau$ with a branching ratio 
(BR) of 85.3\%, while
the only other decay modes are $\chi_2^0 \rightarrow \tilde e_1 e$
and $\chi_2^0 \rightarrow \tilde\mu_1 \mu$. Thus the production
probability for three $\tau$ leptons is high at 85.3\% and the three 
$\tau$-jet
rate is correspondingly 27.2\%.

Example 2 ($\tan \beta = 34$, $\Lambda = 75$\,TeV, $M = 150$\,TeV,
and $n = 1$) and Example~3 ($\tan \beta = 34$, $\Lambda = 85$\,TeV,
$M = 340$\,TeV, and $n = 1$) are similar. 
$\chi_1^\pm \rightarrow \tilde\tau_1 \nu_\tau$ is still very much the
dominant decay mode, but in these cases 
$\chi_1^\pm \rightarrow \chi_1^0 W^\pm$ now
occurs with a small but significant BR\@. On the other hand, the
BR for $\chi_2^0 \rightarrow \tilde\tau_1 \tau$ is closer
to unity at the expense of $\chi_2^0 \rightarrow \tilde e_1 e$
and $\chi_2^0 \rightarrow \tilde\mu_1 \mu$.

\begin{figure}
\begin{center}
\epsfxsize=0.6\textwidth
\epsfbox[83 234 487 548]{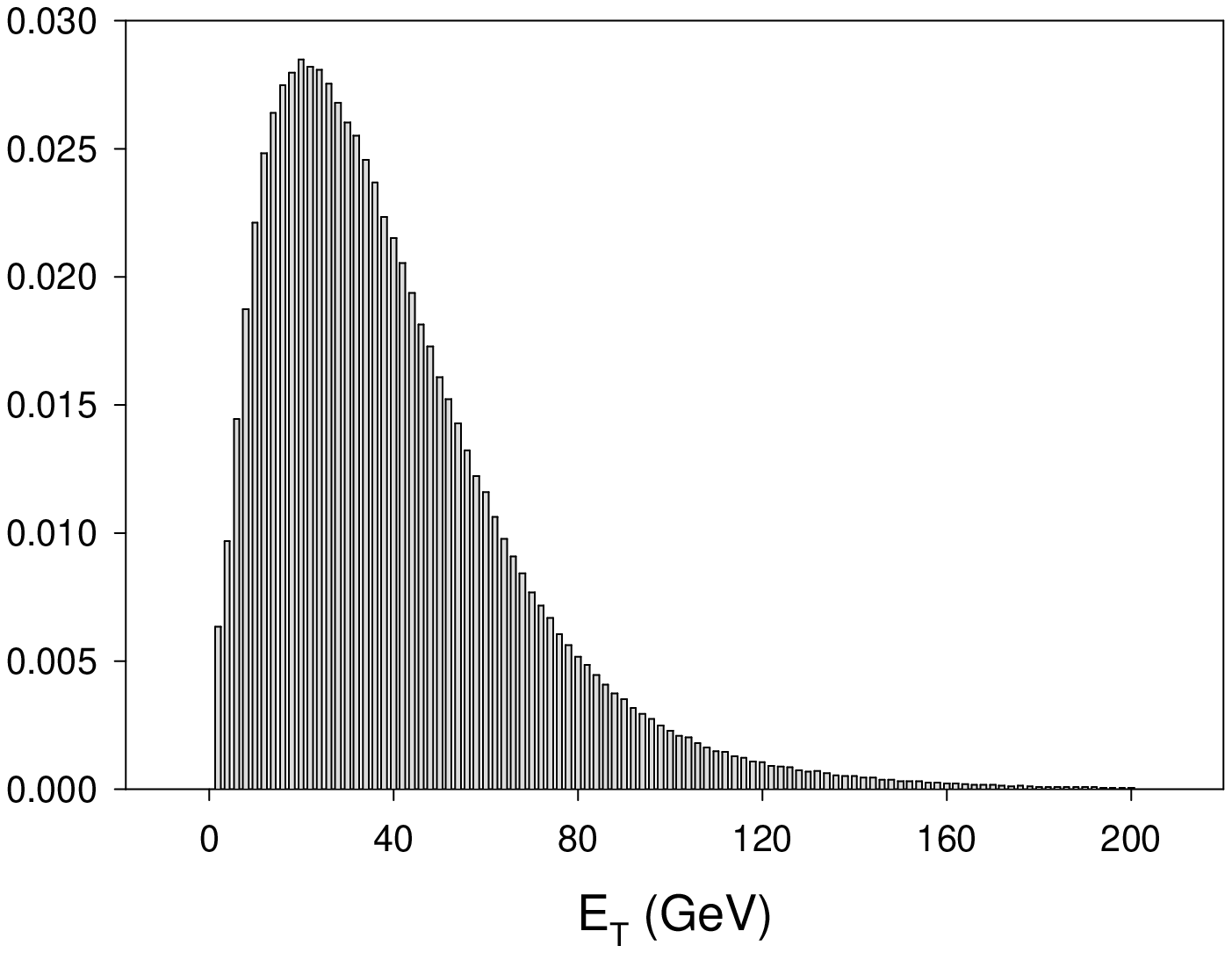}
\begin{minipage}{0.6\textwidth}
(b)
\end{minipage}
\vspace{1.5 cm}
\caption{$E_T$ distribution of the highest $E_T$ $\tau$-jet in 
$\chi_1^+ \chi_1^-$/$\chi_1^\pm \chi_2^0$ production for 
Example~1. 
In (a) no cuts are imposed, while in (b) a pseudorapidity cut of 
$|\eta| < 1$ is imposed on the $\tau$-jets.}
\label{f1l}
\end{center}
\end{figure}

\begin{figure}
\begin{center}
\epsfxsize=0.6\textwidth
\epsfbox[83 234 487 548]{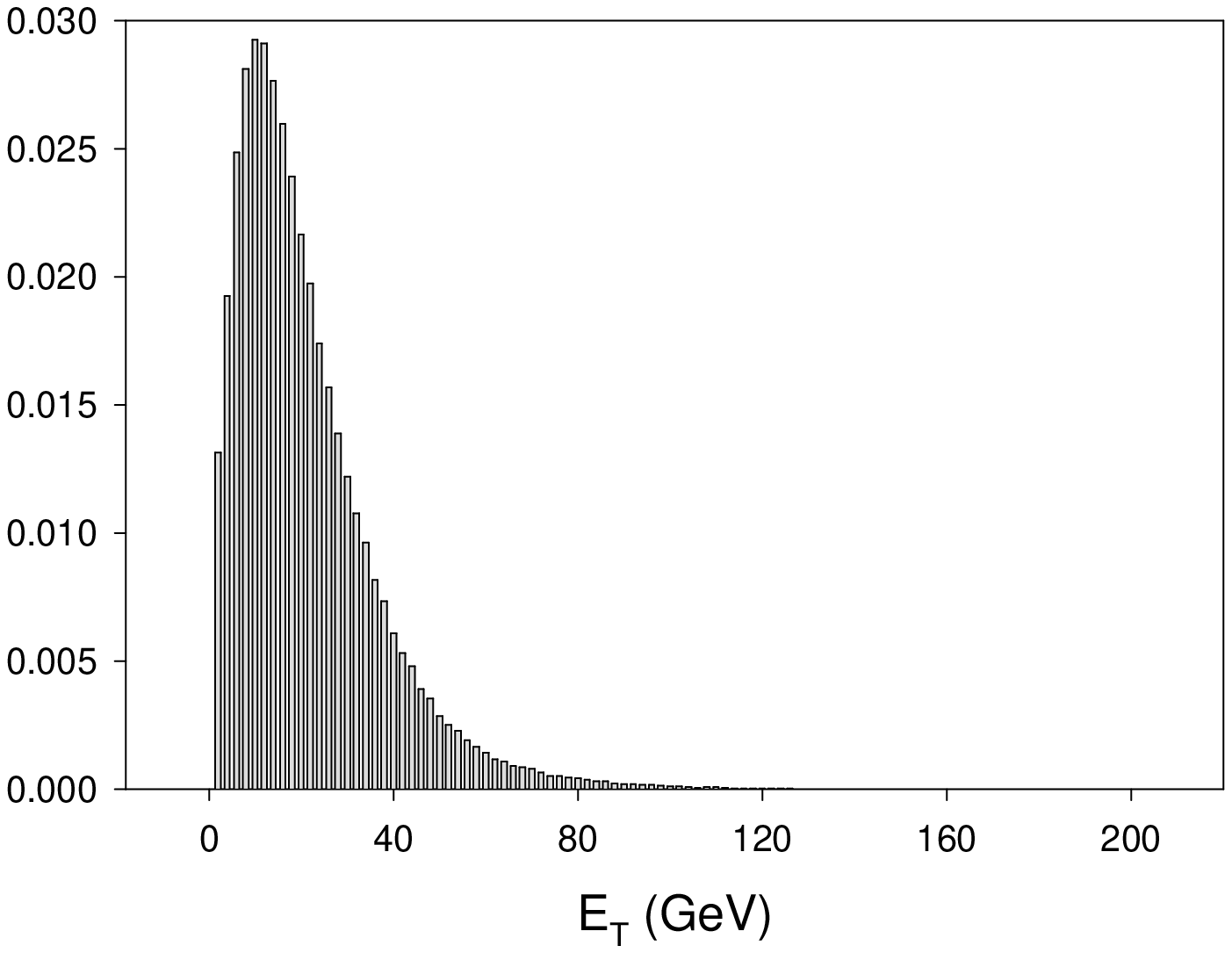}
\begin{minipage}{0.6\textwidth}
(b)
\end{minipage}
\vspace{1.5 cm}
\caption{$E_T$ distribution of the next to highest $E_T$ $\tau$-jet in
$\chi_1^+ \chi_1^-$/$\chi_1^\pm \chi_2^0$ production for Example~1.
In (a) no cuts are imposed, while in (b) a pseudorapidity cut of 
$|\eta| < 1$ is imposed on the $\tau$-jets.}
\label{f1s}
\end{center}
\end{figure}

The question arises as to how high we can expect the $E_T$ of the 
$\tau$-jets to be. Fig.~\ref{f1l} give the $E_T$ distribution of 
the highest 
$E_T$ $\tau$-jet for Example~1. The pseudorapidity cut of $|\eta| < 1$ 
on $\tau$-jets has been  imposed in Fig.~\ref{f1l}(b). 
The peak in the distribution 
occurs at about 
20\,GeV with a broad tail that reaches out beyond 120\,GeV\@. Thus the
leading $\tau$-jets are relatively hard and many will pass the
transverse energy cut of $E_T > 20$\,GeV\@.
The next to 
highest $E_T$ $\tau$-jet is significantly different as Fig.~\ref{f1s} shows.
Here the distribution peaks at a lower value of about 10\,GeV and hardly
extends at all above 80\,GeV\@. 
Due to this softness of the secondary $\tau$-jets, many of the $\tau$-jets
will tend to be eliminated by the cuts.
For Example~1, an $E_T$ cut of 10\,GeV on $\tau$-jets eliminates about 
a third of the second highest $E_T$ $\tau$-jets in those events with
more than 1 $\tau$-jet.

\begin{figure}
\begin{center}
\epsfxsize=0.6\textwidth
\epsfbox[83 234 487 548]{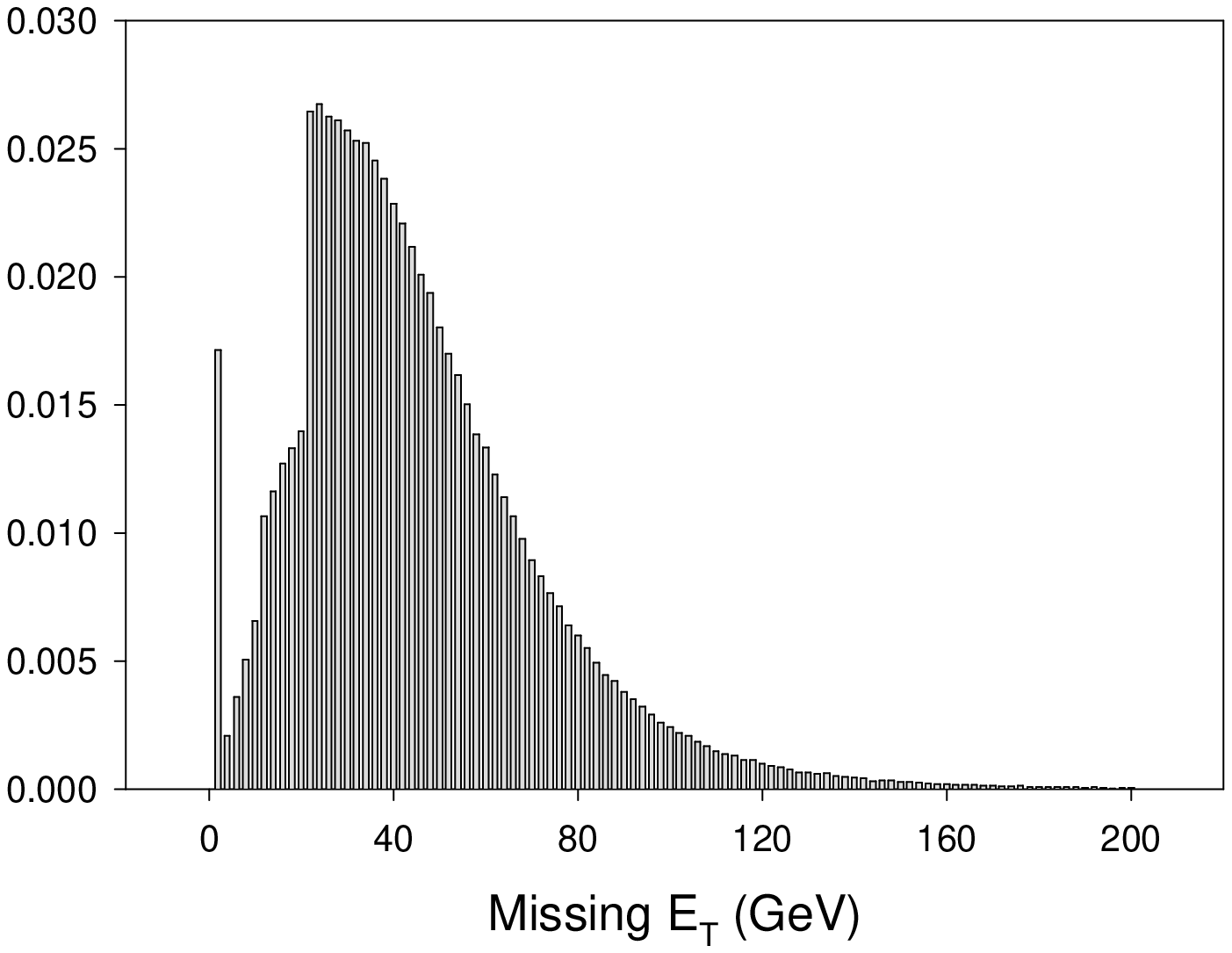}
\begin{minipage}{0.6\textwidth}
(b)
\end{minipage}
\vspace{1.5 cm}
\caption{\met\ distribution for $\chi_1^+ \chi_1^-$/$\chi_1^\pm \chi_2^0$
production for Example~1. In (a) the cuts used are that the jets must 
satisfy $E_T > 10$\,GeV and $|\eta| < 2$ ($|\eta| < 1$ for $\tau$-jets),
while electrons and muons must satisfy $p_T > 10$\,GeV and
$|\eta| < 1$. In (b) the cuts are as in (a) except that the highest 
$E_T$ $\tau$-jet must now satisfy $E_T > 20$\,GeV.}
\label{f1m}
\end{center}
\end{figure}

Also of interest is the \met\ distribution. With energetic and stable 
gravitinos and neutrinos produced in the decays, it is expected that
large missing transverse energy could be an important part of the 
signal. 
Since the missing transverse energy is calculated from what is
observed, however, the question arises as to whether significant
cancellation occurs due to the many decay products.
Fig.~\ref{f1m} gives the \met\ distribution for Example~1. 
The figure demonstrates that the \met\ distribution is indeed broad
with a tail reaching out beyond 120\,GeV\@. The peak occurs at about  25\,GeV
and so a 30\,GeV cut should not be too restrictive.
Examples 2 and 3 have even harder \met\ distributions since their
gaugino masses are larger than for Example~1. Since the \met\ is calculated
from what is observable, the spike at \met\ = 0 is due to events where
none of the jets or charged leptons meet the cuts.

We now consider the specifics of the various final state possibilities.
Table \ref{e1it} gives the inclusive branching ratios for different
numbers of $\tau$-jets for Example~1. 
As indicated above, this example always
produces two $\tau$ leptons in chargino pair production.
Before cuts the inclusive branching
ratio for the 2 $\tau$-jet mode in chargino pair production is 
41.2\%, while the 1 $\tau$-jet inclusive branching ratio is 45.6\%.
After the cuts specified above, the branching ratios are cut down
rather substantially. The one $\tau$-jet BR becomes 22.5\% and
the two $\tau$-jet BR is 9.0\%.
Similar results also hold for the other two cases:
for Example~2, the branching ratios for which are given in Table \ref{e2it},
the before cuts BR for the 2 $\tau$-jet inclusive BR is 42.0\% and the
BR for the 1 $\tau$-jet inclusive BR is 45.0\%. 
There is also a small rate for 3 $\tau$-jets due to one of the $\chi_1^\pm$ 
decaying via the 
$\chi_1^\pm \rightarrow \chi_1^0 W^\pm$ mode followed by the
decay $\chi_1^0 \rightarrow \tilde\tau_1 \tau$.
The cuts don't eliminate
as many of the $\tau$-jets for this case as the lightest chargino is more
massive than for Example 1. After cuts the two $\tau$-jet inclusive
BR is 13\% and the one $\tau$-jet inclusive BR is 29\%. Similar results
hold for Example 3, see Table \ref{e3it}.

$\chi_1^\pm \chi_2^0$ production most frequently produces 3 $\tau$
leptons and thus there is the possibility of having 3 $\tau$-jets in 
these events. For Example~1, the BR before cuts for 3 $\tau$-jets is
27.2\%, while after cuts this goes down to 3.6\%. 
The probability for 2 $\tau$-jets here is larger than for chargino pair
production: before cuts the BR is 44.4\% and after cuts the BR is 17.3\%.
As in the chargino pair case, the reduction after cuts 
for Examples 2 and 3 is less. For example 2,
the inclusive 3 $\tau$-jet BR before cuts is 27.6\% and after the cuts it is
5.7\%.

The question now arises as to the observability of these modes at 
Tevatron's Run II\@. For Example~1, Table \ref{e1it} indicates that the
inclusive 3 $\tau$-jet rate for combined $\chi_1^+ \chi_1^-$
and $\chi_1^\pm \chi_2^0$ production is 9.6\,fb. 
For an integrated
luminosity of 2\,fb$^{-1}$ (approximately what is expected initially 
during Run II), 
this corresponds to $\sim$19 observable events.
For 30\,fb$^{-1}$, the number of observable events is $\sim$288.
The 2 $\tau$-jets cross section of 62.8\,fb gives $\sim$126 events
for 2\,fb$^{-1}$ of data and $\sim$1880 events for 30\,fb$^{-1}$ of
data. For Example~2, the numbers are lower due to the large masses
of the charginos and neutralinos. 
For 2\,fb$^{-1}$ (30\,fb$^{-1}$) of data, the expected number of events
for the 3 $\tau$-jet mode is about 8 (126), while for the 2 $\tau$-jet
mode the expectation is for 49 (737). The number of observed events will 
be less depending on the $\tau$-jet detection efficiency.

The branching ratios for some of the important individual modes are given 
in Table \ref{chgno1}
for chargino pair production
and Table \ref{nlino1} for $\chi_1^\pm \chi_2^0$ production.
The electrons and muons are typically too soft to pass the cuts and thus
requiring an $e$ or $\mu$ to enhance the signal over background
probably will be of little help.

\subsection{Case 2: $M_{\chi_2^0} \approx M_{\chi_1^\pm} > m_{\tilde \nu}
> M_{\chi_1^0} > m_{\tilde e_1, \tilde \mu_1} > m_{\tilde \tau_1}$}

We consider three examples of this case, the masses of
which are given in Table \ref{mass2} and the branching ratios are given
in Table \ref{BR2}.
This case is more complicated than 
the previous one due to the shifting of the sneutrino masses below
that of $\chi_1^\pm$ and $\chi_2^0$ and also to the shifting
of the selectron and smuon masses below the mass of the lightest 
neutralino. As a consequence, there are now many more decay modes for
$\chi_1^\pm$ and $\chi_2^0$. The dominant decay mode of the
lightest chargino is still $\chi_1^\pm \rightarrow  \tilde\tau_1 \nu_\tau$, 
but now the decays to the sneutrinos are also important. 
In fact, the decay to the sneutrinos can have
branching ratios approaching that of the decay to the stau: for
Example 5, which has $\tan \beta = 15$, $M = 400$\,TeV, $\Lambda = 20$\,TeV,
and $n = 4$, we have 
BR($\chi_1^\pm \rightarrow \tilde\tau_1 \nu_\tau$) = 0.279, while
BR($\chi_1^\pm \rightarrow \tilde\nu_\tau \tau$) = 0.237.

For the second lightest neutralino, the dominant
decay mode for these examples is 
$\chi_2^0 \rightarrow \tilde\tau_1 \tau$. But,
as with the decays of the lightest chargino, 
here the decays to the sneutrinos
are important. The branching ratio for the decays of the $\chi_2^0$
to the sneutrinos tend to range from 10\% to 20\% each.

Another important difference in this case from the last one is that 
$\tilde e_1$ and $\tilde\mu_1$ have a lower mass than the lightest
neutralino and therefore can not decay to it. Then the only two-body
decay modes for $\tilde e_1$ and $\tilde \mu_1$ that preserve $R$-parity are 
$\tilde e_1 \rightarrow e \tilde G$ and
$\tilde\mu_1 \rightarrow \mu \tilde G$. Given the smallness
of the coupling involved, there is the possibility that some three-body
decays are important. For $\tilde e_1$ these are
$\tilde e_1^- \rightarrow e^- \tau^- \tilde\tau^+$ and
$\tilde e_1^- \rightarrow e^- \tau^+ \tilde\tau^-$ with corresponding
decays for $\tilde \mu_1$. These three-body decays are dominant for all
three examples studied.

We now consider the details of the various final state possibilities.
Table \ref{e4it} gives the inclusive branching ratios for different
numbers of $\tau$-jets for Example 4. Given the multitude of decay
possibilities presented in Table \ref{BR2}, up to six $\tau$ leptons
can be produced in $\chi_1^+ \chi_1^-$ events. Moreover, virtually all
events produce $\tau$ leptons since the 
$\tilde e_1 \rightarrow e \tilde G$ and 
$\tilde\mu_1 \rightarrow \mu \tilde G$ decays have nearly negligible 
branching ratios at the expense of the $\tau$ producing three-body modes.
Before cuts the inclusive branching ratio for the three $\tau$-jet mode
in chargino pair production is 29.0\%, for the two $\tau$-jet mode it's
33.8\%, and for the one $\tau$-jet mode it's 18.0\%. After the cuts
the branching ratios are cut down rather substantially. The 1
$\tau$-jet BR becomes 25.9\%, the 2 $\tau$-jet BR is 17.0\%, and
the 3 $\tau$-jet BR is 3.7\%.
Similar results also hold for the other two cases: for Example 5,
the branching ratios of which are given in Table \ref{e5it}, the one,
two, and three $\tau$-jet inclusive branching ratios after cuts are
18.7\%, 15.2\%, and 4.8\%, respectively. Similar results also
hold for Example~6 as shown in Table \ref{e6it}.

In principle, up to seven $\tau$ leptons can be produced in 
$\chi_1^\pm \chi_2^0$ events albeit the seven $\tau$ branching ratio
is negligible. The main modes of interest are still the one, two, and
three $\tau$-jet modes. The branching ratios are larger after cuts
for these modes than in the chargino pair case, but the values are 
similar. For Example 4, the branching ratios before cuts for 
3 $\tau$-jets is 33.4\%, while after cuts this is substantially reduced
to 5.1\%. For 2 $\tau$-jets, the BR before cuts for 2 $\tau$-jets is 
33.4\%, while after cuts it is 19.7\%.

\begin{figure}
\begin{center}
\epsfxsize=0.6\textwidth
\epsfbox[83 234 487 548]{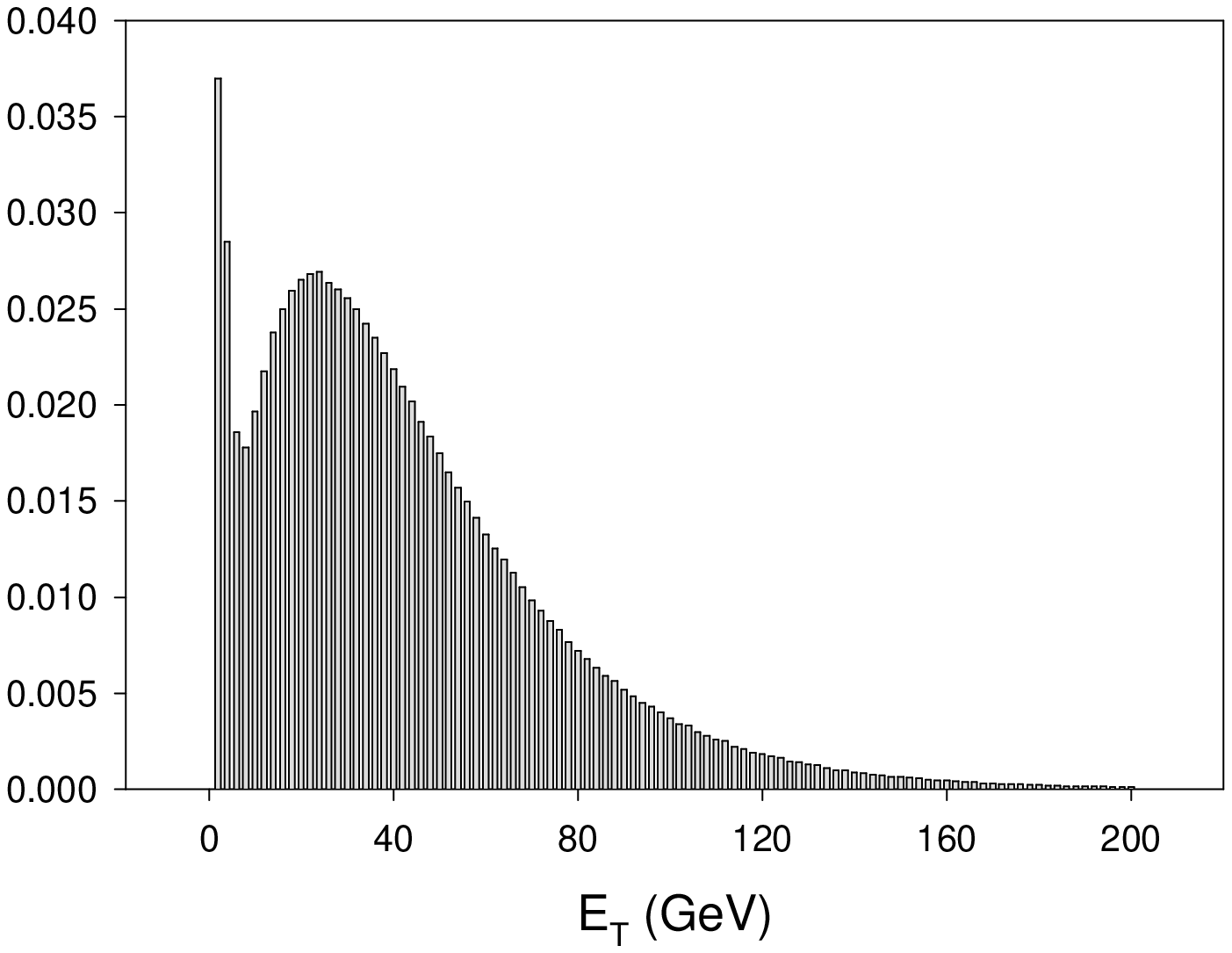}
\begin{minipage}{0.6\textwidth}
(b)
\end{minipage}
\vspace{1.5 cm}
\caption{$E_T$ distribution of the highest $E_T$ $\tau$-jet in 
$\chi_1^+ \chi_1^-$/$\chi_1^\pm \chi_2^0$ production for 
Example~4. 
In (a) no cuts are imposed, while in (b) a pseudorapidity cut of 
$|\eta| < 1$ is imposed on the $\tau$-jets.}
\label{f4l}
\end{center}
\end{figure}

\begin{figure}
\begin{center}
\epsfxsize=0.6\textwidth
\epsfbox[83 234 487 548]{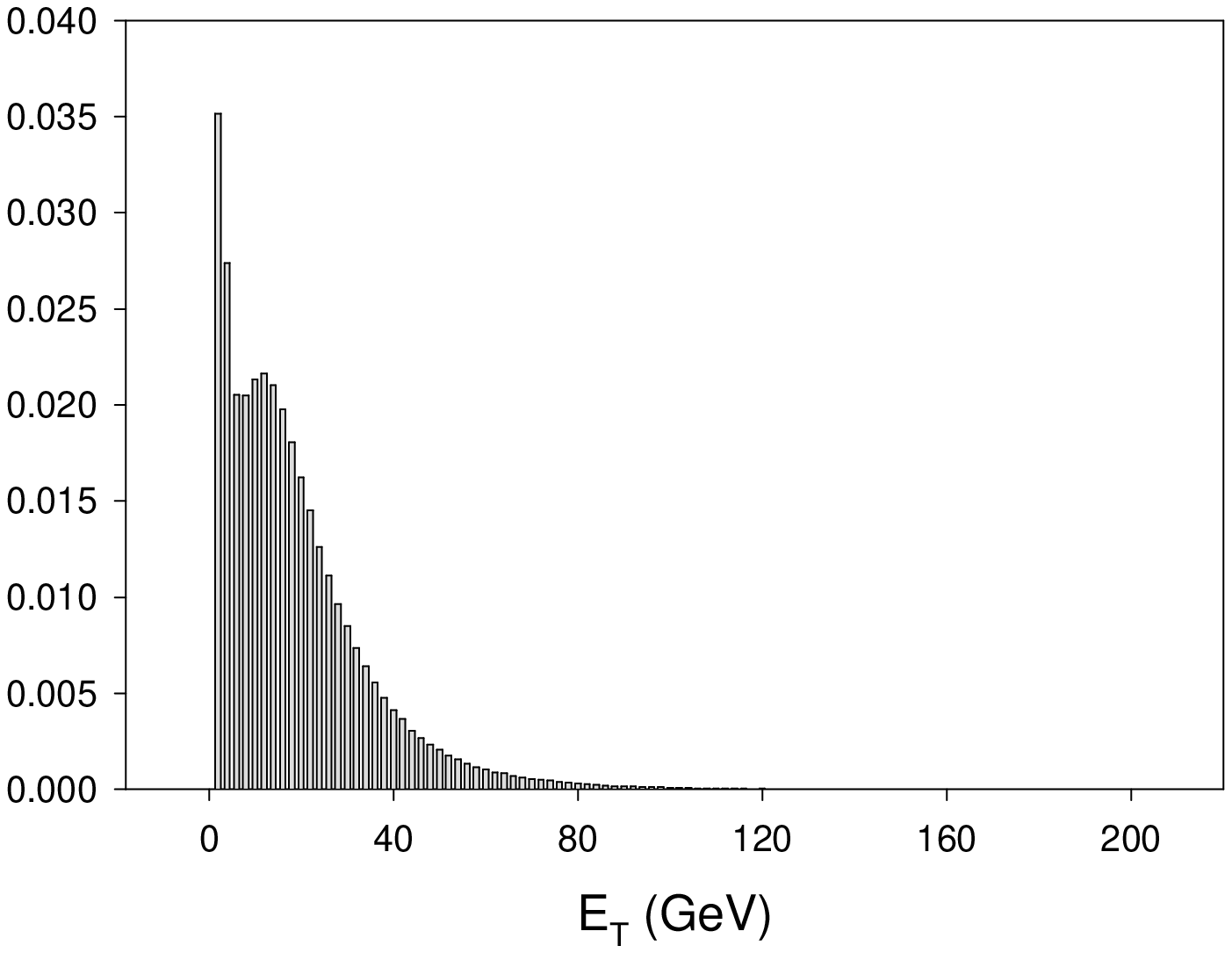}
\begin{minipage}{0.6\textwidth}
(b)
\end{minipage}
\vspace{1.5 cm}
\caption{$E_T$ distribution of the next to highest $E_T$ $\tau$-jet in
$\chi_1^+ \chi_1^-$/$\chi_1^\pm \chi_2^0$ production for Example~4.
In (a) no cuts are imposed, while in (b) a pseudorapidity cut of 
$|\eta| < 1$ is imposed on the $\tau$-jets.}
\label{f4s}
\end{center}
\end{figure}

\begin{figure}
\begin{center}
\epsfxsize=0.6\textwidth
\epsfbox[83 234 487 548]{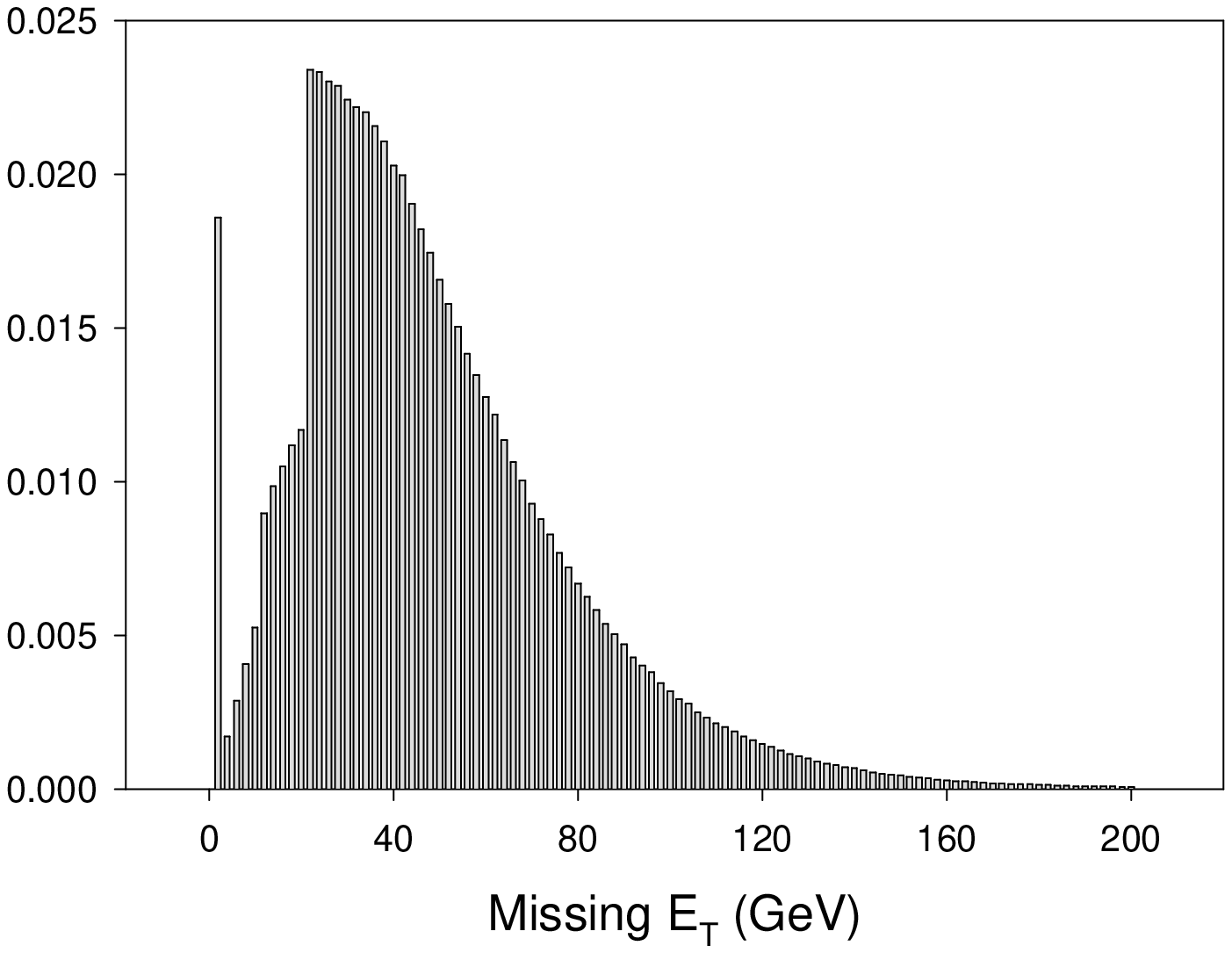}
\begin{minipage}{0.6\textwidth}
(b)
\end{minipage}
\vspace{1.5 cm}
\caption{\met\ distribution for $\chi_1^+ \chi_1^-$/$\chi_1^\pm \chi_2^0$
production for Example~1. In (a) the cuts used are that the jets must 
satisfy $E_T > 10$\,GeV and $|\eta| < 2$ ($|\eta| < 1$ for $\tau$-jets),
while electrons and muons must satisfy $p_T > 10$\,GeV and
$|\eta| < 1$. In (b) the cuts are as in (a) except that the highest 
$E_T$ $\tau$-jet must now satisfy $E_T > 20$\,GeV.}
\label{f4m}
\end{center}
\end{figure}

The $E_T$ distribution of the lead $\tau$-jet for Example~4 is given 
in Fig.~\ref{f4l} and the $E_T$ distribution for the secondary $\tau$-jet
is given in Fig.~\ref{f4s}. 
The distribution for the leading $\tau$-jet is quite
similar to that of Example~1, but the distribution for the secondary 
$\tau$-jet is somewhat softer due to the decrease in the direct
production of $\tau$ leptons from chargino and neutralino decays and
more of the $\tau$ leptons coming from decays further done the decay
chain. The $\met$ distribution of the events are shown in 
Fig.~\ref{f4m}.

What is the probability of observing these events at Tevatron Run II or III? 
For Example~5, the rate for inclusive production of 
3 $\tau$-jets is 8.3\,fb. 
This corresponds to $\sim$16 events for an integrated luminosity
of 2\,fb$^{-1}$ and $\sim$249 events for 30\,fb$^{-1}$.
The inclusive 2 $\tau$-jet rate is 26.2\,fb giving $\sim$52 and
$\sim$786 events for 2\,fb$^{-1}$ and 30\,fb$^{-1}$ of data,
respectively. The values for Example~4 are lower due to the larger 
chargino and neutralino masses. The inclusive 3 $\tau$-jet cross
section is 2.2\,fb and the inclusive 2 $\tau$-jet cross section
is 9.1\,fb. This gives about 4 (66) and 18 (273) events,
respectively.

\section{Conclusion}

We have considered the phenomenology of GMSB models where the lighter
stau is the NLSP and
decays within the detector. 
For this situation, the dominant SUSY production processes at the 
Tevatron are $\chi_1^{+}\chi_1^{-}$ and $\chi_1^{\pm}\chi^0_2$. Their prompt 
decays lead to events containing $2\tau$ or $3\tau$ with high $p_T$ plus 
large missing transverse energy.
These signals are different from the photonic signals 
that have been investigated in GMSB models and the dilepton and trilepton
signals in the usual supergravity models.
Searching for the $\tau$ lepton signals by the hadronic decays of
the $\tau$ leptons to thin jets is complicated by the fact that,
while primary $\tau$-jets can have quite high $E_T$, the secondary
$\tau$-jets tend to be rather soft. As a result, many of the $\tau$-jets
tend to be eliminated by the cuts. Our detailed calculations show that the 
most promising channel
is the inclusive 2 $\tau$-jets channel, although the production of
3 $\tau$-jets can be important at the higher integrated luminosity expected
at Run III\@. The missing transverse energy associated with the events 
is quite large 
providing a good trigger for these events. Good $\tau$ identification will be 
extremely important to detect the signal as well as  a detailed understanding 
of the associated background.

\section*{Acknowledgments}
We thank the members of the GMSB SUSY Run II working group, 
especially S.~Martin and X.~Tata for useful communications. 
Two of us (B.~D. and S.~N.) also thank the Fermilab Theoretical Physics Department, 
especially C.~T.~Hill and J.~Lykken for their warm hospitality and support 
during our summer visit when this work was completed. This work was 
supported by DOE grant numbers DE-FG06-854ER-40224, DE-FG02-94ER 40852 
and DE-FG03-98ER41076.

\begin{table}[p]
\caption{Masses and branching ratios for the three examples of the case where
the ordering of the masses is
$m_{\tilde \nu} > M_{\chi_2^0} \approx 
M_{\chi_1^\pm} > m_{\tilde e_1, \tilde \mu_1} > M_{\chi_1^0} > 
m_{\tilde \tau_1}$.}
\centering
\begin{tabular} {c c c c}
 & Example 1 & Example 2 & Example 3 \\
\hline
 & $\Lambda = 32$\,TeV & $\Lambda = 75$\,TeV & $\Lambda = 85$\,TeV \\
 & $n = 2$, $M = 15\Lambda$ & $n =1$, $M = 2\Lambda$ & $n = 1$, $M = 4\Lambda$ \\
 & $\tan \beta = 20$ & $\tan \beta = 34$ & $\tan \beta = 34$ \\
\hline
$m_h$ (GeV) & 114 & 121 & 122 \\
$m_{H^\pm}$ & 303 & 384 & 451 \\
$m_A$       & 292 & 376 & 444 \\
$m_{\chi_1^0}$ & 83 & 105 & 115 \\
$m_{\chi_2^0}$ & 149 & 195 & 218 \\
$m_{\chi_3^0}$ & -273 & -370 & -439 \\
$m_{\chi_4^0}$ & 299 & 386 & 451 \\
$m_{\chi_{1,2}^\pm}$ & 148, 300 & 194, 388 & 218, 453 \\
$m_{\tilde \tau_{1,2}}$ & 70, 186 & 95, 287 & 113, 323 \\
$m_{\tilde e_{1,2}}$ & 92, 178 & 137, 276 & 156, 313 \\
$m_{\tilde \nu_\tau}$ & 158 & 261 & 300 \\
$m_{\tilde \nu_e}$ & 159 & 264 & 303 \\
$m_{\tilde t_{1,2}}$ & 499, 596 & 794, 883 & 864, 967 \\
$m_{\tilde b_{1,2}}$ & 528, 574 & 818, 885 & 901, 974 \\
$m_{\tilde u_{1,2}}$ & 558, 577 & 870, 903 & 962, 1000 \\
$m_{\tilde d_{1,2}}$ & 557, 582 & 867, 906 & 959, 1004 \\
$m_{\tilde g}$ & 559 & 676 & 737 \\
$\mu$ & -264 & -363 & -433 \\
\hline
$\chi_1^\pm \rightarrow \tilde\tau_1 \nu_\tau$ & 1 & 0.983 & 0.972 \\
$\chi_1^\pm \rightarrow \chi_1^0 W^\pm$ & - & 0.017 & 0.028 \\
$\chi_2^0 \rightarrow \tilde\tau_1 \tau$ & 0.853 & 0.973 & 0.978 \\
$\chi_2^0 \rightarrow \tilde e_1 e$ & 0.073 & 0.013 & 0.008 \\
$\chi_2^0 \rightarrow \chi_1^0 Z$ & - & - & 0.006 \\
$\tilde e_1 \rightarrow \chi_1^0 e$ & 1 & 1 & 1 \\
$\chi_1^0 \rightarrow \tilde\tau_1 \tau$ & 1 & 1 & 1 \\
\hline
$\sigma_{p \bar p \rightarrow \chi_1^+ \chi_1^-}$\,(fb) & 189 & 56.8 & 32.0 \\
$\sigma_{p \bar p \rightarrow \chi_1^\pm \chi_2^0}$\,(fb) & 265 & 73.5 &
39.6 \\
\end{tabular}
\label{mass1}
\end{table}

\clearpage

\begin{table}[p]
\caption{Inclusive branching ratios and production rates for different 
numbers of $\tau$-jets for the case where $\tan \beta = 20$, $M = 480$\,TeV, 
$\Lambda = 32$\,TeV, and $n = 2$. The cross section is for combined 
$\chi_1^+ \chi_1^-$/$\chi_1^\pm \chi_2^0$ production.}
\centering
\begin{tabular}{l c c c}
 & 1 $\tau$-jet & 2 $\tau$-jets & 3 $\tau$-jets \\ 
\hline
$\chi_1^+ \chi_1^-$: w/o cuts & 0.4563 & 0.4120 &  - \\
with cuts     & 0.2249 & 0.0902 & - \\
\hline
$\chi_1^\pm \chi_2^0$: w/o cuts & 0.2406 & 0.4438 & 0.2721 \\
with cuts     & 0.2291 & 0.1726 & 0.0363 \\
\hline
Cross section (fb)  & 103.2 & 62.77 & 9.61 \\
\end{tabular}
\label{e1it}
\end{table}

\begin{table}[p]
\caption{Inclusive branching ratios and production rates for different 
numbers of $\tau$-jets for the case where $\tan \beta = 34$, $M = 150$\,TeV, 
$\Lambda = 75$\,TeV, and $n = 1$. The cross section is for combined
$\chi_1^+ \chi_1^-$/$\chi_1^\pm \chi_2^0$ production.}
\centering
\begin{tabular}{l c c c}
 & 1 $\tau$-jet & 2 $\tau$-jets & 3 $\tau$-jets \\ 
\hline
$\chi_1^+ \chi_1^-$: w/o cuts  & 0.4495 & 0.4197 & 0.00954 \\
with cuts     & 0.2946 & 0.1304 & 0.0005 \\
\hline
$\chi_1^\pm \chi_2^0$: w/o cuts  & 0.2376 & 0.4406 & 0.2756 \\
with cuts     & 0.2789 & 0.2337 & 0.0568 \\
\hline
Cross section (fb)  & 37.22 & 24.58 & 4.20 \\
\end{tabular}
\label{e2it}
\end{table}

\begin{table}[p] 
\caption{Inclusive branching ratios and production rates for different 
numbers of $\tau$-jets for the case where $\tan \beta = 34$, $M = 340$\,TeV, 
$\Lambda = 85$\,TeV, and $n = 1$. The cross section is for combined
$\chi_1^+ \chi_1^-$/$\chi_1^\pm \chi_2^0$ production.}
\centering
\begin{tabular}{l c c c}
 & 1 $\tau$-jet & 2 $\tau$-jets & 3 $\tau$-jets \\ 
\hline
$\chi_1^+ \chi_1^-$: w/o cuts  & 0.4444 & 0.4197 & 0.0159 \\
with cuts     & 0.3286 & 0.1465 & 0.0003 \\
\hline      
$\chi_1^\pm \chi_2^0$: w/o cuts  & 0.2384 & 0.4395 & 0.2734 \\
with cuts     & 0.2979 & 0.2530 & 0.0639 \\
\hline
Cross section (fb)  & 22.30 & 14.70 & 2.54 \\
\end{tabular}
\label{e3it}
\end{table}

\clearpage


\begin{table}[p]
\centering
\caption{Branching ratios for some of the more interesting final state 
configurations with and without cuts in chargino pair production.}
\begin{tabular} {c c c c c c c} 
 & \multicolumn{2}{c}{Example 1} & \multicolumn{2}{c}{Example 2} 
 & \multicolumn{2}{c}{Example 3}\\
 & no cuts & cuts & no cuts & cuts & no cuts & cuts \\
\hline
1 $\tau$-jet            & -      & 0.1303 & -      & 0.1423 & -      & 0.1440 \\
e/$\mu$ + 1 $\tau$-jet  & 0.2281 & 0.0245 & 0.2210 & 0.0400 & 0.2160 & 0.0483 \\
1 jet + 1 $\tau$-jet    & -      & 0.0457 & -      & 0.0651 & -      & 0.0738 \\
2 $\tau$-jets           & 0.4200 & 0.0902 & 0.4055 & 0.1265 & 0.3962 & 0.1398 \\
\end{tabular}
\label{chgno1}
\end{table}

\begin{table}[p]
\centering
\caption{Branching ratios for some of the more interesting final state 
configurations with and without cuts in $\chi_1^\pm \chi_2^0$ production.}
\begin{tabular} {c c c c c c c}
 & \multicolumn{2}{c}{Example 1} & \multicolumn{2}{c}{Example 2} 
 & \multicolumn{2}{c}{Example 3} \\
 & no cuts & cuts & no cuts & cuts & no cuts & cuts \\
\hline
1 $\tau$-jet             & -      & 0.0641 & -      & 0.0620 & -      & 0.0572 \\
e/$\mu$ + $\tau$-jet     & -      & 0.0294 & -      & 0.0364 & -      & 0.0391 \\
2 e/2 $\mu$ + $\tau$-jet & 0.0513 & 0.0051 & 0.0575 & 0.0064 & 0.0578 & 0.0074 \\
e + $\mu$ + $\tau$-jet   & 0.1026 & 0.0069 & 0.1146 & 0.0011 & 0.1151 & 0.0137 \\
2 $\tau$-jets            & -      & 0.0902 & -      & 0.1106 & -      & 0.1119 \\
1 jet + 2 $\tau$-jets    & -      & 0.0328 & -      & 0.0536 & -      & 0.0593 \\
e/$\mu$ + 2 $\tau$-jets  & 0.1894 & 0.0219 & 0.2119 & 0.0319 & 0.2113 & 0.0378 \\
3 $\tau$-jets            & 0.2322 & 0.0344 & 0.2622 & 0.0556 & 0.2573 & 0.0625 \\
\end{tabular}
\label{nlino1}
\end{table}

\clearpage

\begin{table}[p] 
\centering
\caption{Masses and branching ratios for the three examples of the case where
the ordering of the masses is
$M_{\chi_2^0} \approx M_{\chi_1^\pm} > m_{\tilde \nu} > M_{\chi_1^0} > 
m_{\tilde e_1, \tilde \mu_1} > m_{\tilde \tau_1}$.}
\begin{tabular} {c c c c}
 & Example 4 & Example 5 & Example 6 \\
\hline
 & $\Lambda = 32$\,TeV & $\Lambda = 20$\,TeV & $\Lambda = 22$\,TeV \\
 & $n = 3$, $M = 4.69\Lambda$ & $n =4$, $M = 20\Lambda$ & $n = 4$, 
 $M = 40\Lambda$ \\
 & $\tan \beta = 12$ & $\tan \beta = 15$ & $\tan \beta = 18$ \\
\hline
$m_h$ (GeV) & 118 & 115 & 117 \\
$m_{H^\pm}$ & 373 & 301 & 339 \\
$m_A$       & 364 & 290 & 329 \\
$m_{\chi_1^0}$ & 127 & 104 & 116 \\
$m_{\chi_2^0}$ & 224 & 180 & 206 \\
$m_{\chi_3^0}$ & -316 & -268 & -308 \\
$m_{\chi_4^0}$ & 355 & 307 & 341 \\
$m_{\chi_{1,2}^\pm}$ & 222, 355 & 178, 307 & 205, 341 \\
$m_{\tilde \tau_{1,2}}$ & 101, 220 & 71, 174 & 73, 194 \\
$m_{\tilde e_{1,2}}$ & 108, 218 & 86, 168 & 94, 186 \\
$m_{\tilde \nu_\tau}$ & 202 & 147 & 168 \\
$m_{\tilde \nu_e}$ & 202 & 147 & 168 \\
$m_{\tilde t_{1,2}}$ & 658, 764  & 500, 611 & 544, 663 \\
$m_{\tilde b_{1,2}}$ & 710, 735 & 541, 575 &  591, 633 \\
$m_{\tilde u_{1,2}}$ & 726, 748 & 562, 578 & 619, 637 \\
$m_{\tilde d_{1,2}}$ & 725, 752 & 562, 584 & 619, 642 \\
$m_{\tilde g}$ & 826 & 690 & 755 \\
$\mu$ & -310 & -260 & -301 \\
\hline
$\sigma_{p \bar p \rightarrow \chi_1^+ \chi_1^-}$\,(fb) & 22.9 & 71.8 & 37.2 \\
$\sigma_{p \bar p \rightarrow \chi_1^\pm \chi_2^0}$\,(fb) & 26.3 & 89.4 & 45.1 \\
\end{tabular}
\label{mass2}
\end{table}

\clearpage

\begin{table}[p]
\centering
\caption{Branching ratios for the three examples of the case where
the ordering of the masses is
$M_{\chi_2^0} \approx M_{\chi_1^\pm} > m_{\tilde \nu} > M_{\chi_1^0} > 
m_{\tilde e, \tilde \mu} > m_{\tilde \tau_1}$.}
\begin{tabular} {c c c c}
 & Example 4 & Example 5 & Example 6 \\
\hline
 & $\Lambda = 32$\,TeV & $\Lambda = 20$\,TeV & $\Lambda = 22$\,TeV \\
 & $n = 3$, $M = 4.69\Lambda$ & $n =4$, $M = 20\Lambda$ & $n = 4$, 
 $M = 40\Lambda$ \\
 & $\tan \beta = 12$ & $\tan \beta = 15$ & $\tan \beta = 18$ \\
\hline
$\chi_1^\pm \rightarrow \tilde\tau_1 \nu_\tau$ & 0.3397 & 0.2793 & 0.2905 \\
$\chi_1^\pm \rightarrow \tilde\nu_\tau \tau$ & 0.1895 & 0.2371 & 0.2104 \\
$\chi_1^\pm \rightarrow \tilde\nu_e e$ & 0.1777 & 0.2210 & 0.1944 \\
$\chi_1^\pm \rightarrow \chi_1^0 W^\pm$ & 0.1096 & - & 0.0138 \\
$\chi_1^\pm \rightarrow \tilde\tau_2 \nu_\tau$ & - & 0.0025 & 0.0111 \\
$\chi_1^\pm \rightarrow \tilde e_2 \nu_e$ & 0.0064 & 0.0195 & 0.0427 \\
$\chi_2^0 \rightarrow \tilde\tau_1 \tau$       & 0.3824 & 0.3513 & 0.3458 \\
$\chi_2^0 \rightarrow \tilde e_1 e$            & 0.1235 & 0.0812 & 0.0398 \\
$\chi_2^0 \rightarrow \tilde\nu_\tau \nu_\tau$ & 0.1091 & 0.1358 & 0.1450 \\
$\chi_2^0 \rightarrow \tilde\nu_e \nu_e$       & 0.1066 & 0.1332 & 0.1411 \\
$\chi_2^0 \rightarrow \tilde\tau_2 \tau$       & 0.0045 & 0.0096 & 0.0212 \\
$\chi_2^0 \rightarrow \tilde e_2 e$            & 0.0141 & 0.0373 & 0.0632 \\
$\chi_2^0 \rightarrow \chi_1^0 Z$          & 0.0158 & - & - \\
$\chi_1^0 \rightarrow \tilde\tau_1 \tau$        & 0.4573 & 0.5769 & 0.6018 \\
$\chi_1^0 \rightarrow \tilde e_1 e$             & 0.2713 & 0.2115 & 0.1991 \\
$\tilde\nu_\tau \rightarrow \chi_1^0 \nu_\tau$ & 0.8759 & 1 & 0.6775 \\
$\tilde\nu_\tau \rightarrow \tilde\tau_1 W$        & 0.1241 & - & 0.3225 \\
$\tilde\nu_e \rightarrow \chi_1^0 \nu_e$       & 1 & 1 & 1 \\
$\tilde e_2 \rightarrow \chi_1^0 e$            & 1 & 1 & 1 \\
$\tilde e_1 \rightarrow e^- \tau^- \tilde\tau_1^+$ & 0.5874 & 0.6240 & 0.6441 \\
$\tilde e_1 \rightarrow e^- \tau^+ \tilde\tau_1^-$ & 0.4091 & 0.3760 & 0.3559 \\ 
$\tilde e_1 \rightarrow e \tilde G$          & 0.0036 & $\sim10^{-5}$ & $\sim10^{-6}$ \\
$\tilde\tau_2 \rightarrow \chi_1^0 \tau$ & 1 & 1 & 1 \\
\end{tabular}
\label{BR2}
\end{table}

\clearpage

\begin{table}[p]
\centering
\caption{Inclusive branching ratios and production rates for different 
numbers of $\tau$-jets for the case where $\tan \beta = 12$, $M = 150$\,TeV, 
$\Lambda = 32$\,TeV, and $n = 3$. The cross section is for combined
$\chi_1^+ \chi_1^-$/$\chi_1^\pm \chi_2^0$ production.}
\begin{tabular}{l c c c c c}
 & 1 $\tau$-jet & 2 $\tau$-jets & 3 $\tau$-jets & 4 $\tau$-jets &5 $\tau$-jets \\ 
\hline
$\chi_1^+ \chi_1^-$: w/o cuts  & 0.1804 & 0.3378 & 0.2897 & 0.1257 & 0.0282 \\
with cuts     & 0.2529 & 0.1704 & 0.0367 & 0.0040 & 0.0002 \\
\hline
$\chi_1^\pm \chi_2^0$: w/o cuts  & 0.1464 & 0.3344 & 0.3336 & 0.1399 & 0.0216 \\
with cuts     & 0.2687 & 0.1966 & 0.0513 & 0.0048 & 0.0002 \\
\hline
Cross section (fb)  & 12.86  & 9.07   & 2.19   & 0.22   & 0.01   \\
\end{tabular}
\label{e4it}
\end{table}

\begin{table}[p]
\centering
\caption{Inclusive branching ratios and production rates for different 
numbers of $\tau$-jets for the case where $\tan \beta = 15$, $M = 400$\,TeV, 
$\Lambda = 20$\,TeV, and n = 4. The cross section is for combined
$\chi_1^+ \chi_1^-$/$\chi_1^\pm \chi_2^0$ production.}
\begin{tabular}{l c c c c c}
 & 1 $\tau$-jet & 2 $\tau$-jets & 3 $\tau$-jets & 4 $\tau$-jets &5 $\tau$-jets \\ 
\hline
$\chi_1^+ \chi_1^-$: w/o cuts  & 0.1534 & 0.3138 & 0.3058 & 0.1537 & 0.0414 \\
with cuts     & 0.1865 & 0.1522 & 0.0481 & 0.0075 & 0.0005 \\
\hline
$\chi_1^\pm \chi_2^0$: w/o cuts  & 0.1323 & 0.3155 & 0.3391 & 0.1618 & 0.0299 \\
with cuts     & 0.1993 & 0.1709 & 0.0544 & 0.0079 & 0.0003 \\
\hline
Cross section (fb)  & 31.21 & 26.21 & 8.32 & 1.25 & 0.07  \\
\end{tabular}
\label{e5it}
\end{table}

\begin{table}[p]
\centering
\caption{Inclusive branching ratios and production rates for different 
numbers of $\tau$-jets for the case where $\tan \beta = 18$, $M = 880$\,TeV, 
$\Lambda = 22$\,TeV, and $n = 4$. The cross section is for combined
$\chi_1^+ \chi_1^-$/$\chi_1^\pm \chi_2^0$ production.}
\begin{tabular}{l c c c c c}
 & 1 $\tau$-jet & 2 $\tau$-jets & 3 $\tau$-jets & 4 $\tau$-jets &5 $\tau$-jets \\ 
\hline
$\chi_1^+ \chi_1^-$: w/o cuts  & 0.1690 & 0.3354 & 0.3034 & 0.1348 & 0.0252 \\
with cuts     & 0.2112 & 0.1821 & 0.0592 & 0.0089 & 0.0004 \\
\hline
$\chi_1^\pm \chi_2^0$: w/o cuts  & 0.1436 & 0.3286 & 0.3360 & 0.1468 & 0.0203 \\
with cuts     & 0.2173 & 0.1976 & 0.0675 & 0.0110 & 0.0006 \\
\hline
Cross section (fb)  & 17.65 & 15.68 & 5.24 & 0.82 & 0.04  \\
\end{tabular}
\label{e6it}
\end{table}

\newpage


\begin{figure}
\begin{center}
\epsfxsize=0.6\textwidth
\epsfbox[83 234 487 548]{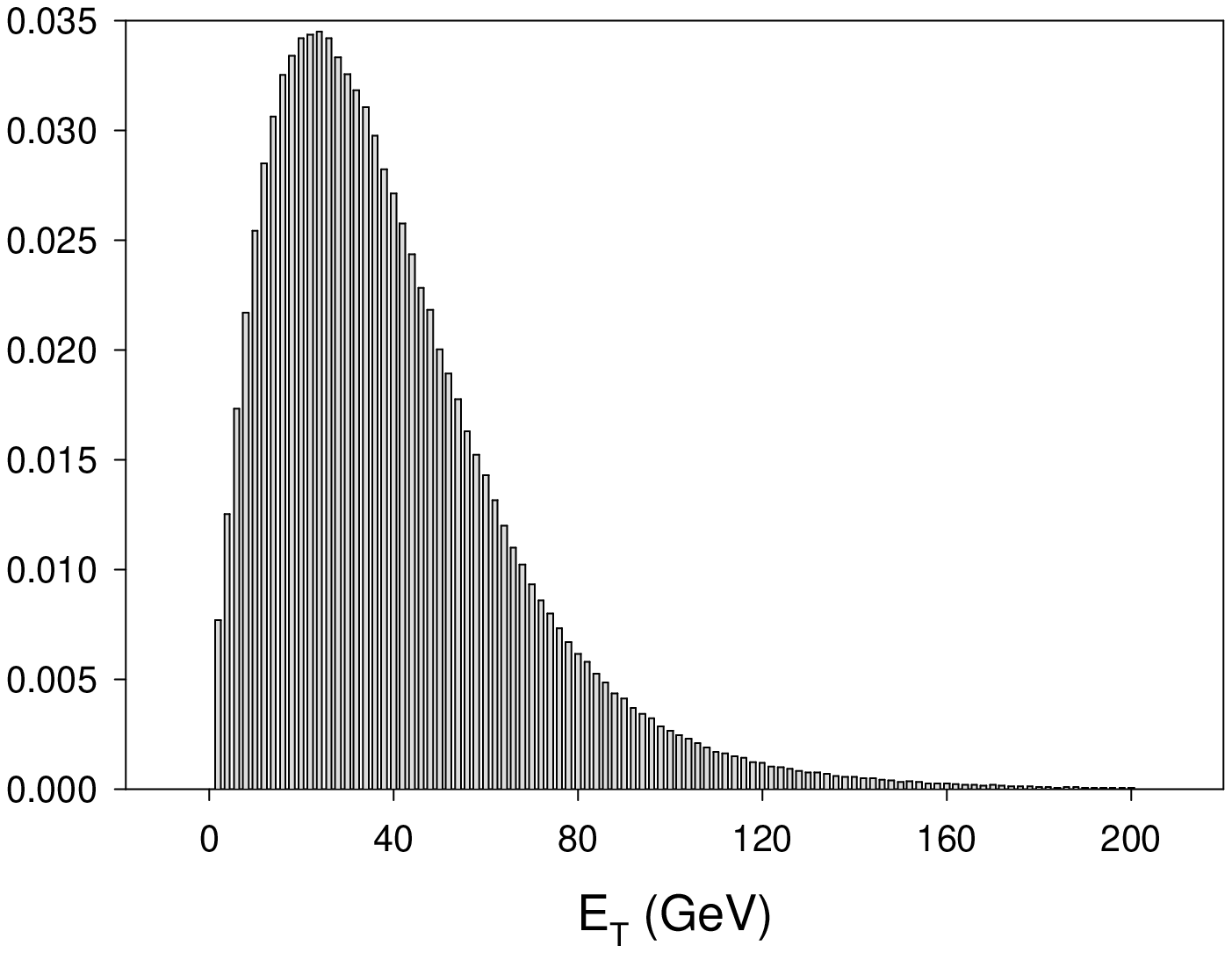}
\begin{minipage}{0.6\textwidth}  
(a)
\end{minipage}
\end{center}
\end{figure}

\vspace{0.5 cm}

\clearpage

\begin{figure}
\begin{center}
\epsfxsize=0.6\textwidth
\epsfbox[83 234 487 548]{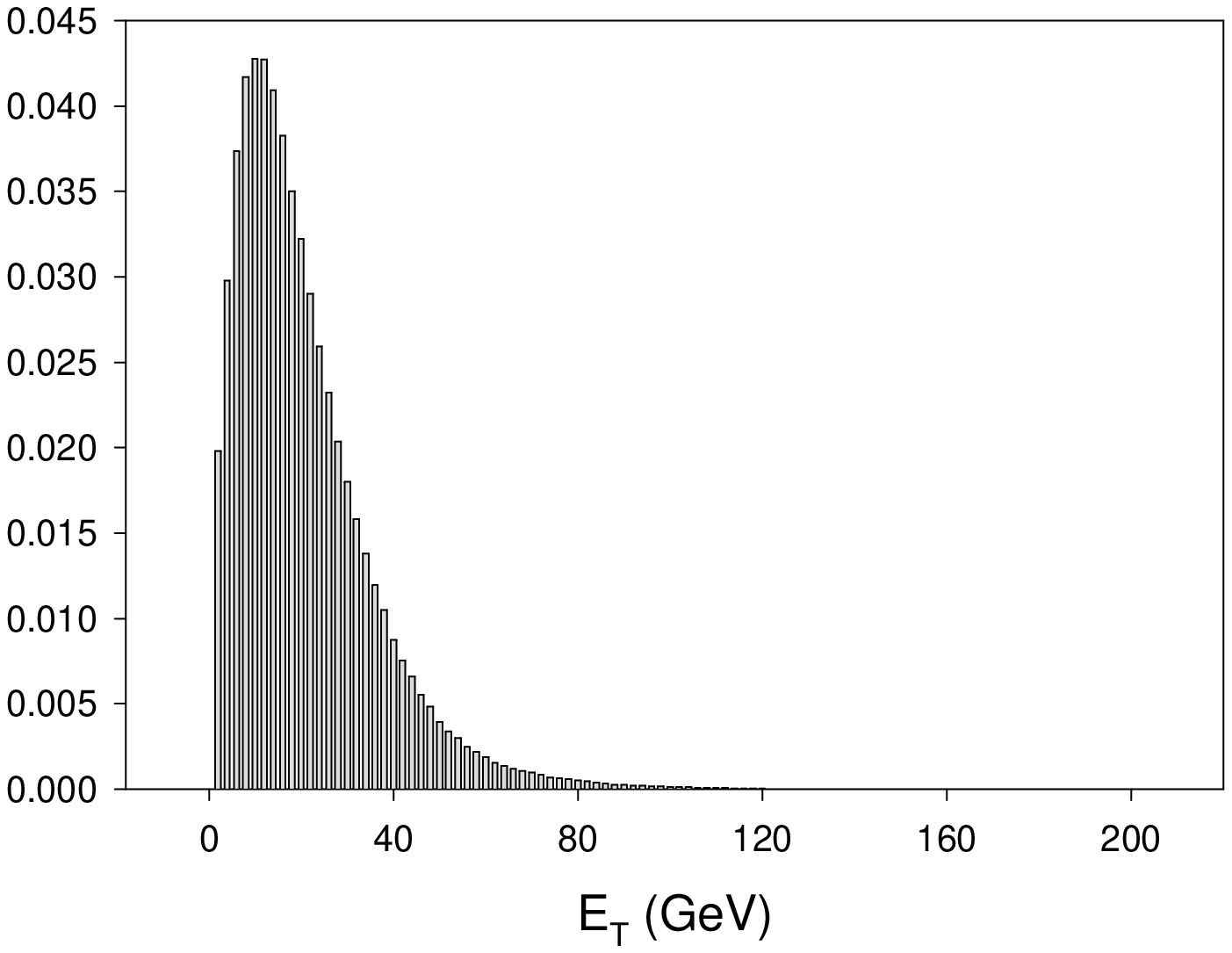}
\begin{minipage}{0.6\textwidth}  
(a)
\end{minipage}
\end{center}
\end{figure}

\vspace{0.5 cm}

\clearpage

\begin{figure}
\begin{center}
\epsfxsize=0.6\textwidth
\epsfbox[83 234 487 548]{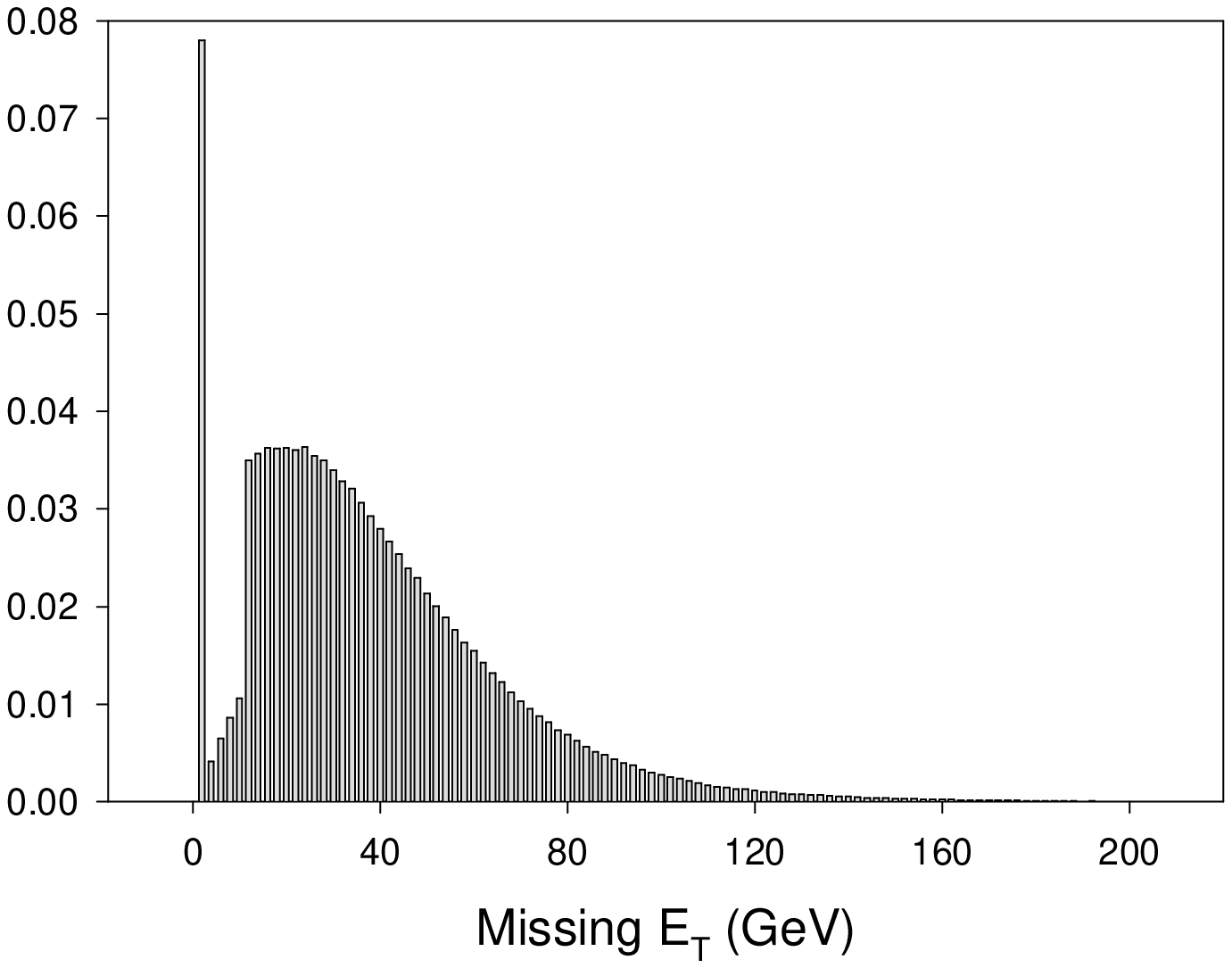}
\begin{minipage}{0.6\textwidth}  
(a)
\end{minipage}
\end{center}
\end{figure}

\vspace{0.5 cm}

\clearpage


\begin{figure}
\begin{center}
\epsfxsize=0.6\textwidth    
\epsfbox[83 234 487 548]{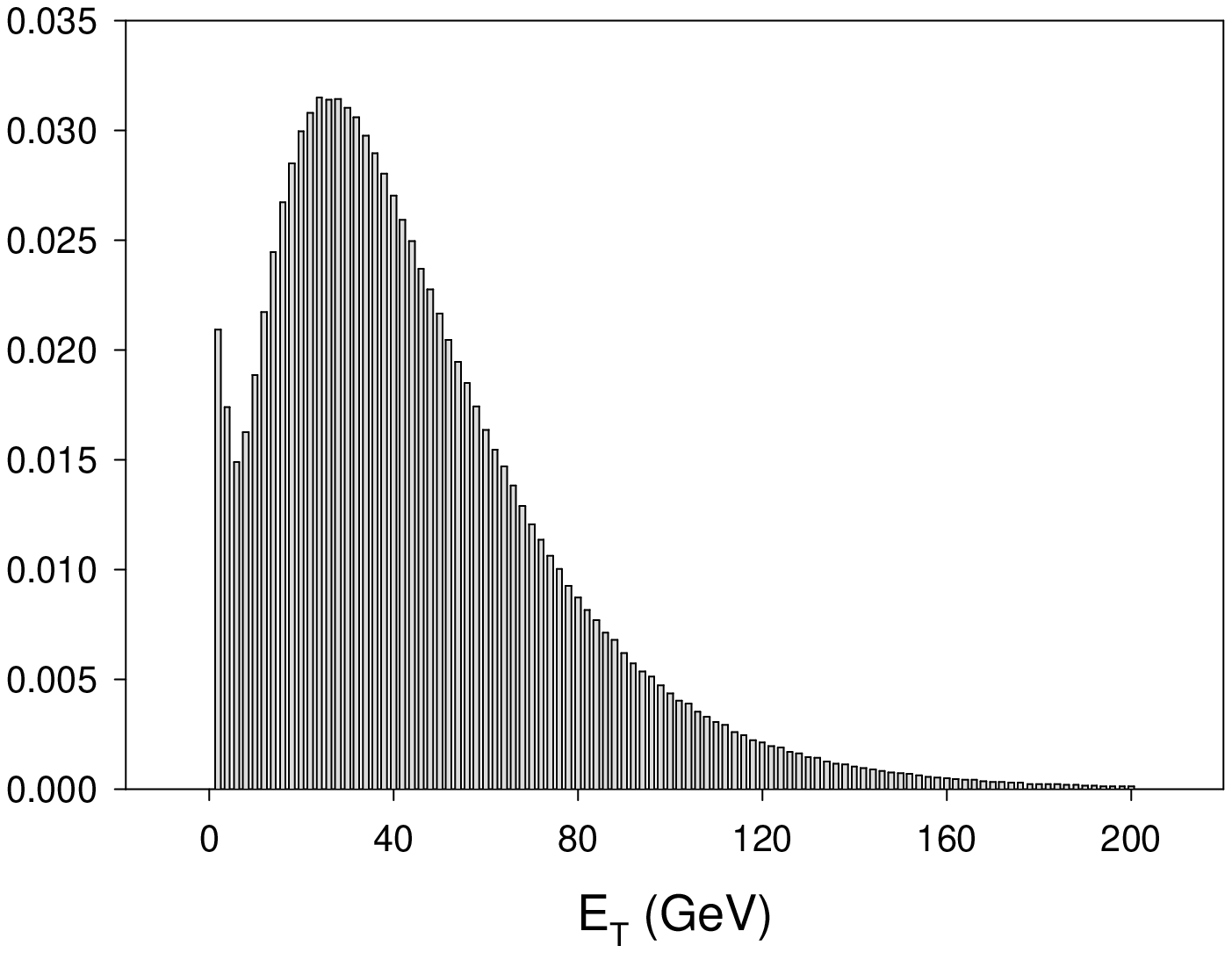}
\begin{minipage}{0.6\textwidth}  
(a)
\end{minipage}
\end{center}
\end{figure}

\vspace{0.5 cm}

\clearpage

\begin{figure}
\begin{center}
\epsfxsize=0.6\textwidth
\epsfbox[83 234 487 548]{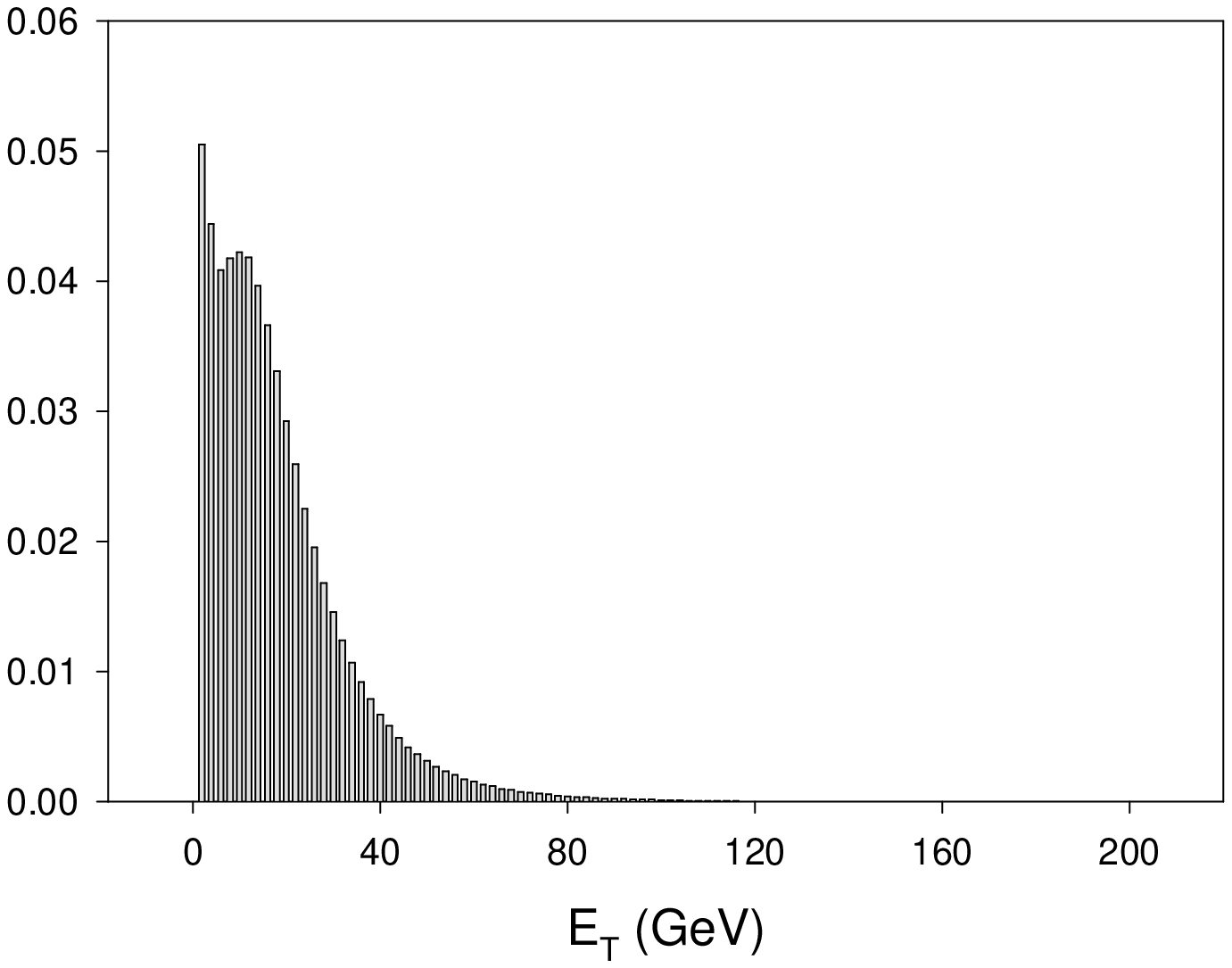}
\begin{minipage}{0.6\textwidth}  
(a)
\end{minipage}
\end{center}
\end{figure}

\vspace{0.5 cm}

\clearpage

\begin{figure}
\begin{center}
\epsfxsize=0.6\textwidth
\epsfbox[83 234 487 548]{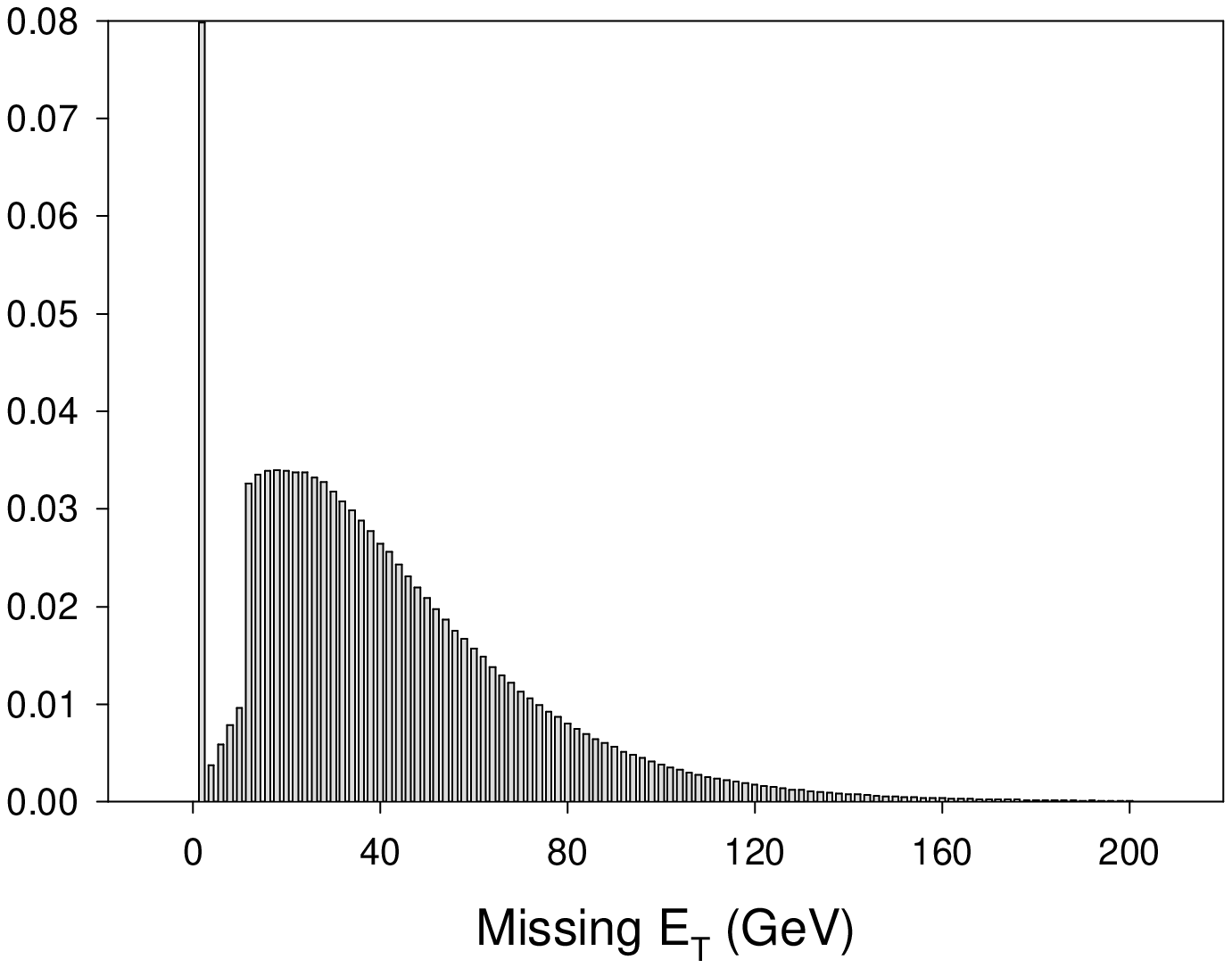}
\begin{minipage}{0.6\textwidth}  
(a)
\end{minipage}
\end{center}
\end{figure}

\vspace{0.5 cm}

\end{document}